# Responsible AI by Design in Practice

**Richard Benjamins, Alberto Barbado, Daniel Sierra**

Telefónica, Ronda de la Comunicación, 28050 Madrid, Spain
{richard.benjamins, alberto.barbadogonzalez, daniel.sierraramos}@telefonica.com

## Abstract

Recently, a lot of attention has been given to undesired consequences of Artificial Intelligence (AI), such as unfair bias leading to discrimination, or the lack of explanations of the results of AI systems. There are several important questions to answer before AI can be deployed at scale in our businesses and societies. Most of these issues are being discussed by experts and the wider communities, and it seems there is broad consensus on where they come from. There is, however, less consensus on, and experience with how to practically deal with those issues in organizations that develop and use AI, both from a technical and organizational perspective. In this paper, we discuss the practical case of a large organization that is putting in place a company-wide methodology to minimize the risk of undesired consequences of AI. We hope that other organizations can learn from this and that our experience contributes to making the best of AI while minimizing its risks.

## Introduction

Artificial Intelligence (AI) is on the rise. It can be applied to many different domains such as content recommendations, chatbots, image recognition, machine translation, fraud detection, medical diagnosis, autonomous vehicles, legal, education, transport, and logistics to name just a few. It can not only be used for business, but also for social purposes such as better understanding and reducing the impact of climate change, natural disasters, and migration (Globalpulse 2017).

However, recently several concerns have been expressed about the use of AI, in particular related to potential discrimination (bias, discrimination, predictive parity) (O'Neill 2016), interpretability of algorithmic conclusions (explainability, black box problem) (Samek, Wiegand, Müller 2017, Guidotti et al 2018, Pedreschi et al 2018) transparency of data used (Gross-Brown 2015), impact on jobs (Manyika and Sneader 2018) liability questions (Kingston 2016), and malicious use of the technology (Pistono and Yampolskiy, 2016).

With the increasing uptake of AI technologies in organizations, many start to worry about those concerns, and are wondering how to prepare themselves to avoid unintended negative consequences. While some of the concerns are outside the scope of private enterprises and require governments to act (such as impact on jobs, liability and malicious use), others need to be addressed at the individual company level.

In this paper, we will describe how a large organization as Telefonica (24 countries, 127.000 employees, 343 million accesses, 52 billion in revenues) is approaching this problem. In the next section we briefly describe related work. Then we present the AI principles that Telefonica has developed for responsible AI. Next, we present the "Responsible AI by Design" methodology for implementing the Principles in our organization. After that, we present our experience with the implementation of the methodology in our organization. Then we summarize the challenges we encountered as well as some topics that need further research. Finally, we present a conclusion.

## Related work

Several works have been published about the ethical and societal impact of AI. (Anderson and Anderson 2007) describe how machine behavior is ethically acceptable by human users. (Boden et al 2017) proposes a set of principles to be included in a potential regulation for the design of robots. (Bryson and Winfield 2017) propose the use of standardization such that AI and Autonomous systems exhibit the required algorithmic transparency. (Dignum 2018, 2017) distinguishes between Ethics *by / in / for* Design, where *ethics in design* refers to "ensuring that development processes are aligned with ethical principles." (Gumbus and Grodzinsky 2015) review the relation of AI algorithms with undesired discrimination and relate that to privacy of people in a HR context of organizations.

However, few publications exist on how companies should go about the creation of AI systems while respecting the ethical and societal implications they might have. The work that comes closest to this is IBM's work on AI and Ethics (IBM 2018a). The European Commission has published Ethics guidelines for trustworthy AI (EC 2018) proposing an assessment check list for AI practitioners based on seven principles: Human agency and oversight; technical robustness and safety; privacy and data governance; transparency, diversity, non-discrimination and fairness; societal and environmental wellbeing; accountability.

The principles we are presenting are broadly in line with the EC principles, even though the EC published them six months later. The main differences stem from:

- Our principles focus on the aspects that make AI different from other technologies unless they play a key role for AI such as privacy and security, in which case we refer to existing company practices rather than defining things again. The EC guidelines are more inclusive of all aspects involved in end-to-end AI.
- The EC guidelines aim to cover any sector, whereas our principles- while broader than only for telecoms- do not cover aspects clearly out of the telecom scope. This means for instance, that there is no AI principle for "safety" or "robustness". Of course, safety and robustness are key elements in our operations (e.g. of mobile antennas), but this is not specifically related to AI.
- As discussed in this paper, our principles are part of a methodology which includes a governance model. Some aspects of the EC guidelines are captured in this governance model, such as "accountability".

**Principles of Artificial Intelligence**

As many businesses, Telefonica is strongly committed to respecting Human Rights, as is stated in its Business Principles and Human Rights Policy. This includes a commitment to developing products and services aimed at making the world a better place to live and mitigating any negative impacts technology may have on society or the environment. Technology should contribute to making society more inclusive and offer better opportunities for all, and AI can contribute to these goals.

In order to guide the organization in its uptake of AI and Data across the business, Telefonica has published its "Principles of AI" (Telefonica 2018). The principles include:

- **Fair AI** seeks to ensure that the applications of AI technology lead to fair results. This means that they should not lead to discriminatory impacts on people (Stoyanovich, Abiteboul, Miklau 2016) in relation to race, ethnic origin, religion, gender, sexual orientation, disability or any other personal condition. When optimizing a machine learning algorithm, we must take into account not only the performance in terms of error optimization, but also the impact of the algorithm in the specific domain.
- **Transparent and Explainable AI** means to be explicit about the kind of personal and/or non-personal data the AI systems uses as well as about the purpose the data is used for. When people directly interact with an AI system, it should be clear to the users that this is the case. When AI systems take, or support, decisions, a certain level of understanding of how the conclusions are arrived at needs to be ensured (Samek, Wiegand, Müller 2017), by generation explanations about how they reached that decision, like is illustrated in (Gilpin et al., 2019) for the particular case of supervised machine learning. Those explanations should always consider the user profile to adjust them to the transparency level required, as stated in (Theodorou, Wortham, Bryson, 2017). This also applies in case of using third-party AI technology.
- **Human-centric AI** means that AI should be at the service of society and generate tangible benefits for people. AI systems should always stay under human control and be driven by value-based considerations. AI used in products and services should in no way lead to a negative impact on human rights or the achievement of the UN's Sustainable Development Goals.
- **Privacy and Security by Design** means that when creating AI systems, which are fueled by data, privacy and security aspects are an inherent part of the system's lifecycle. This maximizes respecting people's right to privacy and their personal data. Notice that the data used in AI systems can be personal or anonymous/aggregated. Notice also that this principle is broader applicable than only to AI systems, and most organizations already have processes in place to ensure proper privacy and security.

The Principles are by extension also applicable when working with partners and third parties.

Those Principles are based on a broad consensus in expert communities that have sparked profound conversations and discussions about the impact of AI in society. One just has to read the national strategies of different countries on AI. There are other concerns about AI such as the impact of jobs, liability, malicious use or AI for warfare, but these fall outside the acting scope of private organizations and are in the realm of governments and institutions.

It is one thing to come up with a set of guiding principles for how to design, develop and use AI in organizations, but it is another thing how to implement those principles across organizations such that they have the desired effect. While there is significant consensus on the concerns of AI underlying the principles, there is less experience on how this can be practically implemented.

# Towards a methodology for "Responsible AI by Design"

We have taken the principles a step further and have turned them into a methodology for creating Responsible AI by Design, in the tradition of Privacy and Security by Design.

The methodology consists of the following ingredients
- The AI principles setting the values and boundaries
- A set of questions and check points, ensuring that all AI principles have been considered in the creation process
- Tools that help answering some of the questions, and help mitigating any problems identified
- Training, both technical and non-technical
- A governance model assigning responsibilities and accountabilities

Designing the methodology requires a cross-enterprise initiative involving different departments such as Engineering, Security, Legal, Business, IT, Human Resources, Procurement, as well as an endorsement of top management.

Table 1 below illustrates how the Principles are implemented in the organization. For each principle, several questions are defined that must be answered by the respective responsible persons. Several of the questions require certain understanding of AI and Machine Learning, and therefore specific tools and training are required. The next sections will indicate the details and proposals for each of the ingredients mentioned before.

Table 1: Questions and their corresponding proposals.

| Principle | Question to be asked | Implemented through |
|---|---|---|
| Fair AI | Does your data set contain sensitive variables? | Training |
| | Does any of the variables strongly correlate with sensitive variables? | Technical tool |
| | Is/are your training data set(s) biased with respect to the target groups in case those include "protected groups"? | Technical tool |
| | Is there an important impact in the specific domain of false positives (FP) and/or false negatives (FN)? | Training |
| | Are FP and FN unequally distributed across different (protected) groups | Technical tool |
| Transparent & Explainable AI | Could the user think that s/he interacts with a person rather than with your system? | Training |
| | Is the AI system's outcome significantly affecting people's lives? | Training |
| | Do you lack sufficient understanding of how the AI-generated decisions are constructed for the domain at hand? | Training |
| | Could the user request an explanation for the AI-generated conclusion? | Training |
| | Is it difficult to be explicit about whether the data used is personal or non-personal, and about the purpose the AI system uses the data for? | Training |
| | Is it possible to understand how the algorithm has reached its conclusions? For example, what variables have influenced the result of the algorithm and how much? | Technical tool |
| Human-centric AI | Is there a possibility that your P&S has a negative impact on Human Rights? | Training |
| | Does your P&S negatively impact the UN's SDGs? | Training |
| Privacy & Security by Design | Does your AI system use personal data? | Training |
| | Has your Privacy Impact Assessment revealed any important concerns? | Training |
| | In case your P&S uses anonymized data, is there an unreasonable risk of re-indentification? | Technical tool |
| | Has your Security Assessment revealed any important concerns? | Training |
| | Is the system robust against attacks that seek to exploit weaknesses in it and manipulate the outputs? | Technical tool |
| Third parties | Do you need more information from your supplier to understand whether the AI module is consistent with the Principles? | Training |

## Training content

Given how new the use of AI is in organizations, specific new training material needs to be developed for answering questions related to Transparent & Explainable AI. Some of the questions are not specific to AI and we can refer to existing training material in the organization. This includes the Principles related to Human-centric AI and Privacy & Security by Design. For some questions such as those related to "human-centric AI", "Privacy" and "third-parties" it is possible to refer to existing templates (ADAPT 2017, ICO, Hind 2018) that can be reused for training purposes and for the actual work as well. In Table 2, we show the essential content of the different training components to be developed. Ideally, to ensure "explainable" AI, technical tools would be needed, in addition to training, to support developers.

Table 2: Key content of the training components for each of the AI principles and questions.

| Principle | Question to be asked | Training content |
|---|---|---|
| Fair AI | Does your data set contain sensitive variables? | Explain what is sensitive personal data as defined by law |
| | Is there an important impact of false positives (FP) and/or false negatives (FN)? | Show examples of FP and FN and their concrete impact on people |
| Transparent & Explainable AI | Could the user think that s/he interacts with a person rather than with your system? | Show examples of people and machines interacting with different degrees of machine sophistication |
| | Is the AI system's outcome significantly affecting people's lives? | Show examples of different degrees of decisions' impact on peoples' lives |
| | Do you lack sufficient understanding of how the AI-generated decisions are constructed for the domain at hand? | Explain what types of algorithms are explainable (e.g. decision trees) and what not (e.g. deep learning). Show different ways of how AI decisions can be explained (algorithm, outcome) |
| | Could the user request an explanation for the AI-generated conclusion? | This may require building in certain features during design |
| | Is it difficult to be explicit about whether the data used is personal or non-personal, and about the purpose the AI system uses the data for? | Explain what is personal data, psydonymized data and anonymous data. Give examples of how certain types of data are used for different purposes in AI systems |
| Human-centric AI | Is there a possibility that your P&S has a negative impact on Human Rights? | General training on Human Rights and examples of P&S that impact them in both positive and negative ways. See Ethics Canvas (ADAPT 2017) |
| | Does your P&S negatively impact the UN's SDGs? | Explain the 17 SDGs of the UN and give examples of P&S that impact them in both positive and negative ways |
| Privacy & Security by Design | Does your AI system use personal data? | Explain what is personal data, psydonymized data and anonymous data. |
| | Has your Privacy Impact Assessment revealed any important concerns? | Explain Privacy by Design, and what a PIA is, using for examples templates from ICO (ICO) |
| | Has your Security Assessment revealed any important concerns? | Explain Security by Design |
| Third parties | Do you need more information from your supplier to understand the AI module is consistent with the Principles? | Explain what kind of questions to ask a provider and how to interpret the answers (Hind et al 2018) |

## Technical tools required

Many of the tools required to support "Responsible AI by Design" are still in early stage, though they are being developed quickly and its expected that the companies include them soon in their processes. The technical tools proposal that appears within this methodology spans from the usage of stand-alone tools to the reference of more academic research solutions in order to offer a complete portfolio that could be useful to all the profiles within the company.

Table 3 Summarizes the key functionalities of the required tools.

Table 3: Types of technical tools to help visualize and detect potential problems.

| Principle | Question to be asked | Type of tool |
|---|---|---|
| Fair AI | Does any of the variables strongly correlate with sensitive variables? | Check correlations between all variables in data set and visualize result |
| | Is/are your training data set(s) biased with respect to the target groups in case those include "protected groups"? | Check for disparate impact, i.e. whether different subgroups are treated differently |
| | Are FP and FN unequally distributed across different (protected) groups | Check for predictive parity, i.e. what is the false positive rate for different subgroups as well as for the overall population |
| Transparent & Explainable AI | Is it possible to understand how the algorithm has reached its conclusions? For example, what variables have influenced the result of the algorithm and how much? | Define what transparency level is required according to the profile of the users of the system and use an available solution to generate local or global explanations |
| Privacy & Security by Design | In case your P&S uses anonymized data, is there an unreasonable risk of re-identification? | Notify when anonymization algorithm is not sufficient to render data sets anonymous |
| | Is the system robust against attacks that seek to exploit weaknesses in it and manipulate the outputs? | Before passing a model to production, it is recommended to assess its vulnerabilities to these attacks with the tools available |

**Fair AI**

Regarding Fair AI, some of solutions exist for internal use such as Facebook's Fairness Flow tool (CNET 2018), while others such as IBM's Fairness 360 toolkit (IBM 2018b) and Accenture's AI Fairness tool (Accenture 2018) are for any organization to use. Some tools are open source such as IBM's tool, a tool from Pymetrics (2018), and a tool from the University of Chicago (Aequitas 2018). As these tools focus on "fair AI" their mission is helping to detect and avoid undesired discrimination through bias. Typical functionalities include: detecting bias in data sets related to sensitive data (impacting protected groups), detecting correlations in data sets between normal variables and sensitive variables, detecting unbalanced outcomes of algorithms within sub groups of the population and mitigating the effect of the bias.

The concept of Fair AI has obtained a lot of relevance in the last years. Many new studies are published every year dealing with the problem unfair AI and how to mitigate the inclusion of new biases in the decision-making process. There is some confusion about why bias arises and even if the bias is something natural that must be preserved in order to maintain the "purity" of original data. The potential causes of bias were first registered by (Selbst, Barocas 2016). They listed some causes why a machine learning model can give unfair results

- Skewed data: Incorrect assumptions about the data generation process. It occurs when the data acquisition process is biased.
- Tainted data: Incorrect problem definition and target labeling.
- Limited features: When the number of features is so limited that bias is induced in some sensitive attributes
- Sample size disparities: Unequal sample sizes of different sensitive groups
- Proxy features: Presence of correlated variables within the problem that induces bias even when the sensitive features are removed.

There are several definitions of Fairness provided in the literature: unawareness, group fairness, individual fairness and counterfactual fairness. The definition of fairness through unawareness consists of removing the sensitive variable from input data. This can be insufficient because the presence of proxy features implicitly maintains the information of the deleted sensitive variable. Group fairness deals with Fairness from the perspective of all individuals, while individual fairness tries to model the differences between each subject with the rest of population. The case of counterfactual fairness goes one step beyond trying to interpret the causes of bias via causal graphs.

(Hardt et al. 2016) propose a framework for group fairness in which three different criteria can be employed to evaluate a supervised ML model in terms of Fairness:

- Independence: It is achieved when the model prediction is independent of the sensitive variable, that is, the proportion of Positive samples given by the model is the same for all sensitive groups.
- Separation: Also known as Equalized Odds. It is achieved when the model prediction is independent of the sensitive variable given the target variable, that is, when the TP (true positive) rate and the FP (false positive) rate are equal in all sensitive groups, respectively.
- Sufficiency: Also known as Predictive Rate Parity. It is achieved when the target variable is independent of the sensitive attribute given the model output, that is, when the Positive Predictive Value is the same in all sensitive groups.

It is impossible to achieve all three criteria at the same time, but they can be optimized jointly in order to optimally mitigate bias in ML models.

In terms of Fairness in ML, two main actions can be considered: evaluation and mitigation. The former is the process of measuring and quantifying the amount of bias present in the model (in terms of one or several criteria), and the latter is the process of fixing some aspects of the model in order to reduce or remove the effect of bias in terms of one or several sensitive attributes.

For evaluation, several metrics have been proposed in the last years for the different fairness criteria. In the case of the Independence criterion, possible metrics include statistical parity difference or disparate impact for the case of independence. For the Separation criterion, possible metrics include equal opportunity difference and the average odds difference (Hardt et al. 2016). Another metric is the Theil index (Speicher el al. 2018) which measures the inequality not only in terms of the group fairness, but also in terms of the individual fairness.

For the mitigation phase, several techniques can be found in the literature. In this part, there are three main groups of techniques.

- Pre-processing: These techniques are applied before the machine learning algorithm is trained in

- order to remove biases in the very early stage of the learning process.
- In-processing: These techniques are applied during the training process by including Fairness optimizations constraints along with cost functions in ML models.
- Post-processing: Employed after the algorithm is built, these are the less intrusive techniques because they don't modify the input data or the ML algorithm. This technique is especially adequate for mitigating biases in models that already exist.

Each technique applies the mitigation process in different phases of a typical analytics pipeline and the choice must be made to fit the particular case, but in terms of performance it is better to apply pre-processing or in-processing techniques.

A very simple pre-processing technique is Reweighing (Kamiran and Calders, 2012) consisting of modifying the weights of samples in order to remove discrimination in sensitive attributes. Another example is the technique proposed by (Zemel et al. 2013) in which they transform input data in order to find a good representation that obfuscates information about the membership in the sensitive groups.

Another popular algorithm to mitigate bias is Adversarial Debiasing (Zhang et al. 2018). This in-processing technique tries to maximize the ability of predicting the target variable while minimizing the ability of predicting sensitive variables through GAN.

Also, the case of Equalized Odds post-processing (Hardt et al. 2016) is a good example of mitigation through a post-processing technique. In their proposal, they try to adjust the thresholds in a classification model in order to reduce the differences between the True Positive Rate and False Positive Rate for each sensitive group.

**Transparent & Explainable AI (xAI)**
Explainable AI is something that has existed in the literature for many years. Expert Systems were among the first AI systems that dealt with it, trying to use some of the multiples rules generated as a base for the explanations provided. This was the case of MYCIN (one of the first Expert Systems for diagnosing meningitis, Clancey 1983) which was able to give the following types of explanations: *why was a given fact used? Why was a given fact not used? How was a given conclusion reached? How was it that another conclusion was not reached?* The conclusions were explained in a human-understandable manner by tracing back from the conclusions to the initial state.

In the context of Machine Learning, there are several proposals for xAI. Supervised ML models can be classified into two groups: white box models and black box models. On the one hand, white box models, such as simple decision trees or linear regressions, are models that can generate comprehensive explanations based on the model itself. On the other hand, black box models, such as complex deep learning (DL) architectures, can't provide direct explanations for the decision taken by the system in a way that is comprehensive for a human being. The main issue is that there is a tradeoff between complexity and explainability: more complex models can potentially be more precise, but in exchange the model is opaquer. To be able to use more complex models while being able to generate explanations that can be understood, there are different proposals available depending on the explanations desired.

One example is explaining why a data point is classified within one of the categories used by the supervised model. For that purpose, there are libraries such as LIME (Ribeiro, Singh, Guestrin, 2016) that do not need information about the model itself (model-agnostic). Other proposals like Layer-wise Relevance Propagation (LRP) (Bach et al., 2015) obtain, for deep learning models, the contributions of the different input features to a classification by aggregating the contributions of each of the layers present. This second group of solutions are known as model specific. (Samek, Wiegand, Müller 2017) provide several examples of applying LRP to image recognition. A system trained with pixels of an image for classifying it as containing a "a cat" or "not a cat", will highlight the relevant pixels associated to the category chosen. Then, a person can see the conclusion (e.g. a cat) and can inspect the highlighted pixels to see if they really corresponded to the pixels of the cat (and thus, the system can be trusted). Otherwise, if they correspond to something not related to a cat, even if the accuracy of the algorithm is high, the system should not be trusted. This approach can be applied to a variety of domains and data structures (tabular, texts...).

There are many situations when the desired explanation is not about an individual classification but about how the whole model works and what the main features are that have contributed to the training phase. To address that, there are several solutions, such as surrogate models implemented through libraries such as Skater (Oracle 2017). This solution uses a white box model such as a decision tree trained with the predicted outputs of the black box model (instead of the real values) while using the same input features. This trained white box model then serves as a simple but useful way to understand the most relevant features of the black box model, regardless of the complexity of the model itself. See (Guidotti et al 2018) for a survey of methods for explaining black box models.

Though most of the research work addresses supervised learning, there are also proposals in the scientific literature for other kind of models such as for reinforcement learning (Anderson et al., 2019). Finally, some xAI tools developed by big tech companies are What-If (Google, 2018) and Interpret (Microsoft, 2019).

**Privacy & Security by Design**
Regarding the Privacy & Security by design principle, an example of a tool that contributes to privacy by testing the risk of re-identification in case the data set is anonymized, even if differential privacy techniques are used, is (Dwork, Roth, 2014).

Related to security, one aspect to consider is if the system is robust against attacks that seek to exploit weaknesses and manipulate its outputs. There are techniques, such as Adversarial Examples (Goodfellow, Shlens, Szegedy, 2014), that seek to cheat and manipulate the outputs of a model. In the case of a supervised ML, this takes place by checking the minimum changes in the input data that would cause different classifications. This has happened, for example, with computer vision systems of autonomous vehicles; with slight changes in a stop sign, which go unnoticed by the human eye, the systems detected them instead as speed limit signs of 45 mph (Eykholt et al., 2018).

Some tools available to test and even secure the system against those attacks are for example Cleverhans (Papernot et al, 2016) for deep learning models, AlfaSVMLib (Xiao et al., 2015) for SVM models, AdversarialLib for evasion attacks (Biggio, Corona et al., 2013), and even for unsupervised learning such as clustering algorithms (Biggio, Pillai et al., 2013).

## Implementation of the methodology

The implementation of the methodology is no different from implementing other technology-related policies such as Security & Privacy by Design, but given how unknown the innerworkings of AI are for the wider audience, it might cause additional challenges.

Firstly, the company needs to start an **awareness** campaign explaining what AI is, why and how it is used in the company, and what the challenges are. The campaign introduces the Principles, the methodology, the training program and the tools.

Secondly, the **training** program starts with the people closest to designing and developing products & services that use AI, including the procurement people who deal with buying technology from third parties. In a later phase, training can be extended more widely across the organization, but always starting with people who are most likely to get involved in AI-related initiatives. Training ranges from very technical training to non-technical training, for instance non-technical training for call center or social media agents who might receive customer inquiries related to AI. We have developed an online course of about one hour where we explain to employees basic AI concepts such as Machine Learning, the potential undesired impact of AI such as bias, the AI principles, technical tools and questionnaires, the governance model, and several practical examples of assessing products against the principles. Depending on the role employees have, they will see more or less technical details.

As discussed (see Table 1 and Table 3), several **specific tools** are needed to support the implementation of Responsible AI in an organization. Without those tools, it is very hard to make justifiable statements about fairness or explainability.

We believe that as a start, self-control is more effective than control through specific committees. Therefore, we start with an agile **governance** model, delegating as much as possible responsibility to the people responsible for the concerned products and services, and providing them with support in case a question is raising concern. Making people aware of the concerns and providing them with concrete training and tools will help changing the culture to creating Responsible AI by Design. In this way, we manage to "sensibilize" the organization without imposing a lot of controls. Such awareness is also important for potential future regulation. The governance model only covers the part that is specific for AI. For privacy and security matters, existing governance models are used.

## Challenges and research needs

### Fairness

One of the main challenges around Fair AI is the evangelization of users, developers and companies about the need for Fairness treatment in ML processes. Most people believe that bias is something intrinsic to the nature of process that generates data and must not be altered. As an example, if we examine the proportion of men and women studying engineering, we will find that, in general, women are more inclined to other type of studies. We can think that people chose whatever they want (and it's true) and data do not suffer from any type of human bias, but the truth is that all decisions are strongly affected by society and stereotypes that determine the shape of data in the acquisition process.

Fair AI has been one of the most important topics in the last years. Many studies and new techniques have arisen around how to make fair models. The constant increasing of work and new perspectives about the problem has created a need for a unified framework for Fairness in AI in order to simplify the process of evangelization and implementation, especially in the industry.

Several techniques as disparate impact or Equal Opportunity difference are intended to measure specific aspects of Fairness in ML models. Other metrics like Theil index are useful for group fairness and individual fairness but in general, the need of a unified metric for this purpose is more evident every day. With the inclusion of unified technique of Fairness, the process of evaluation and correction would be easier to implement and compare.

### Explainability

Though not every domain is equal, and explainability is more relevant for applications such as medical diagnosis than for recommendations of movies, most domains, including telecommunications, require some kind of explainability. The type of explanation should be adapted to the different profiles and transparency levels required by the domain and, sometimes, regulations. For example, according to GDPR, Article 13 users have the right to know "meaningful information about the logic involved" (GDPR, 2018) when personal data is used, and thus, implying that explainability is compulsory for any organization delivering products and services in Europe in that scenario.

Besides the explicit requirements within regulation, xAI could be provided for all profiles that have a relationship with the AI system, including technical profiles (Data Scientists, DevOps...), stakeholders (in case of private companies), domain experts (such as doctors for medical AI applications), regulators and auditors, and end users of the system (like patients in case of medical applications). This poses a challenge for xAI since most of the aforementioned tools assist in providing a fixed type of explanation, whereas ideally explanations should be tailored to the level of transparency required and to the specific profile at hand. It would be even better if the explanations generated, would take into account specific domain knowledge, such that they go beyond mere rankings of features in the case of supervised ML.

There are some research proposals that deal with this, such as (Hind, Wei, Campbell, Codella, 2019) where a supervised model is trained, not only with the labels to predict, but also with the domain explanations they should provide. In this approach domain explanations are linked to particular combinations of feature values and their corresponding prediction. Any time the model predicts a new instance, it will include an explanation. This approach can be used to train a model with the adequate explanations for a transparency level, like, for instance, the transparency level required for an end user. As a consequence, the user will receive explanations according to the data points, but the inner information of the model itself will be still be unknown to them, and thus, it offers a privacy-conserving explainability method.

Another proposal tackling the same problem is presented in (Doran, Schulz, Besold, 2017) where the relevant features detected in a supervised ML model are combined with a knowledge-based reasoning system in order to provide human comprehensible explanations.

While those proposals are promising, they are still early-stage research, and therefore hard to automate for inclusion in production environments.

Another important challenge regarding xAI, mentioned by (Guidotti et al 2018), is that there is still not an available formalism to define a common reference for what an explanation should be. There are some criteria to consider while evaluating an explanation (it should be concise, consistent, should consider anomalous situations…) but that is still not enough to set a general reference to build them. Also, the article mentions that explanations are still constrained to the available features of a data set (for supervised ML) but they do not consider other relevant features such as latent variables, correlated information that is not explicitly present in the dataset, etc... So, building explanations based only on those features might be too simplistic.

Other recent proposals are related to the definition of methodologies and procedures to document and standardize the development of an AI software product in order to reflect how company AI principles have been addressed. (Hind et al., 2018) suggests creating documentation, inspired by the SDoCs (supplier's declaration of conformity, which describes the lineage of a product including its safety and performance tests), which contain the purpose, performance, safety, security and the origin of information of an AI product, and that can be examined by clients. This serves as a mechanism for the builder of an AI system or service to communicate how all relevant issues have been addressed in the product or service.

**Methodology**

A challenge for a methodology for Responsible AI by Design is the trade-off between innovation speed and risks. The speed of technological changes and their uptake in the market dictate a fast incorporation of AI in products and services. However, risk-reducing methodologies usually slow down the innovation pace due to controls by committees. That is the reason why we have chosen to start with self-control by using the methodology as an awareness and training tool for our employees. It may delay slightly the development and go-to-market processes, but the likelihood of social acceptance of the solutions is expected to be higher, and the risk for committing an error with negative impacts, will be reduced.

## Use Cases at Telefónica

Currently, at Telefónica, there are some products and uses cases that address the AI principles of the company.

Regarding fairness, Telefónica developed a tool coined Luca Ethics that allows to audit and mitigate bias in any existing classification ML model. The methodology has two main phases: the first one consists of quantifying the bias present in the model and the second one consists of applying post-processing techniques to mitigate the bias. For bias measurement, several state-of-the-art techniques are employed like statistical parity, average odds difference, predictive parity difference or mutual information. Once the bias is quantified, we optimize the classification threshold of each sensitive group individually in order to satisfy the equal opportunity criterion (Hardt et al. 2016) while achieving the best possible model performance.

Regarding transparency and data privacy, Telefónica has created a "transparency center" for users of Aura, Telefonica's virtual assistant. This center allows customers to have direct access to, and control over their personal data held by the company and used by the system. Another product related to privacy is Spectra, a software tool for complex and robust data anonymization, along with a patented tool to avoid the risk of re-identifying anonymized data.

In the area of xAI, as an example we mention two products that include an explainability feature: A device recommender for mobile handsets, and Luca Comms for analyzing organizational communications patterns. Device Recommender is a software product that recommends a device to a specific user using techniques of reinforcement learning (contextual bandits) combined with market research information elicited by a conjoint analysis. The recommender system includes an automatic natural-language generated explanation that incorporates the key features of why that device was considered. The second example, Luca Comms, is a big data analytics service that identifies opportunities for business improvement within the complexity of enterprise communication data. Luca Comms includes a patent-pending (Barbado et al., 2019) xAI solution for unsupervised ML anomaly detection models (OneClass SVM) for local explanations (why a particular data point is anomalous) using model specific information and providing counterfactual explanations through a visual interface.

# Conclusions

In this paper, we have discussed a methodology called "Responsible AI by Design" that can be used by organizations who plan to use AI at a large scale, and want to avoid unintentionally creating undesired side-effects such as unfair discrimination, or not being able to understand the conclusions of an AI system. The methodology starts with defining a set of overall AI principles that state the values and boundaries. Other ingredients include awareness & training, a set of questions to be asked in the development process, specific tools to help answering some of the questions, and a governance process defining responsibilities and accountability. It is the combination of those different elements that makes the methodology successful. With the expected massive uptake of AI, we think that such methodologies are becoming increasingly important, and practical experience needs to be shared to ensure the sustainability of AI. Like many other organizations, we have only started the journey. We do therefore not consider our principles and methodology as final standards for ethical AI, but as a starting point that will evolve as more collective experience becomes available.